\documentclass[twocolumn,showpacs,amsmath,pra,aps,amssymb,superscriptaddress,nofootinbib]{revtex4-1} 
\usepackage[T1]{fontenc}
\usepackage[latin9]{inputenc}
\usepackage{times}
\usepackage{color} 
\usepackage{xspace}
\usepackage{amssymb,amsmath}
\usepackage{amsbsy}
\usepackage[pdftex]{graphicx}
\usepackage{bm}
\usepackage{float}
\usepackage{adjustbox}
\usepackage{epstopdf}

\usepackage[unicode,breaklinks]{hyperref}
\hypersetup{
    unicode=true,
    plainpages=false, 
    colorlinks=true,
    linkcolor=blue,
    citecolor=blue,
    filecolor=black,
    urlcolor=blue
}
\usepackage{url}
\usepackage{verbatim}

\begin{document}

\title{Dynamics of polar-core spin vortices in a ferromagnetic spin-1 Bose-Einstein condensate}
\author{Lewis A. Williamson}  
\affiliation{Dodd-Walls Centre for Photonic and Quantum Technologies, Department of Physics, University of Otago, Dunedin 9016, New Zealand}
\author{P.~B.~Blakie}  
\affiliation{Dodd-Walls Centre for Photonic and Quantum Technologies, Department of Physics, University of Otago, Dunedin 9016, New Zealand}

\begin{abstract}
A ferromagnetic spin-1 condensate supports polar-core spin vortices (PCVs) in the easy plane phase. We derive a model for the dynamics of these PCVs using a variational Lagrangian approach. The PCVs behave as massive charged particles interacting under the two dimensional Coulomb interaction, with the mass arising from interaction effects within the vortex core. We compare this model to numerical simulations of the spin-1 Gross-Pitaevskii equations and find semi-quantitative agreement. In addition, the numerical results suggest that the PCV core couples to spin waves, and this affects the PCV dynamics even far from the core. We identify areas of further research that could extend the model of PCV dynamics presented here.
\end{abstract}

\maketitle

\section{Introduction}
Spinor Bose-Einstein condensates can exist in various ferromagnetic or antiferromagnetic phases depending on the nature of the spin dependent interactions and Zeeman shifts of the spin levels. Associated with this range of symmetry breaking phases is a rich array of topological defects~\cite{StamperKurn2013a,Kawaguchi2012R}. In a ferromagnetic spin-1 condensate, the Zeeman shifts arising from an external field can be tuned so that the condensate magnetizes in a plane perpendicular to the external field. This inplane transverse magnetization breaks the $\text{U}(1)$ rotational symmetry about the direction of the external field. This phase is termed the \emph{easy plane} phase and can support polar-core spin vortices (PCVs). The transverse magnetization angle exhibits a phase winding around a PCV, with a filled but unmagnetized core. The easy plane symmetry breaking and PCVs have been observed \emph{in situ} experimentally in a spin-1 condensate~\cite{Sadler2006a}. PCVs form spontaneously after a quench to the easy plane phase, and are intimately linked with Kibble-Zurek effects~\cite{Saito2007a,Uhlmann2007a,Lamacraft2007a} and post-quench phase ordering dynamics~\cite{Saito2007b,williamson2016a,williamson2016b}. PCVs may also play a role in spin turbulence~\cite{fujimoto2012,fujimoto2014} and, at non-zero temperature, could possibly give rise to a Berezinskii-Kosterlitz-Thouless transition to the easy plane phase.  Despite the relevance of PCVs to a wide variety of interesting physical processes, a comprehensive study of their properties is lacking.

A model of the dynamics of PCVs was presented by Turner in~\cite{turner2009}, where hydrodynamic arguments were used to show that the PCVs should interact like massive charged particles according to the two dimensional Coulomb's law. The presence of a logarithmic Coulomb interaction is known in scalar vortices~\cite{pismen1999,kardar2007}, however the presence of the vortex mass makes Turner's dynamic model of PCVs vastly different from scalar vortex dynamics. Although a hydrodynamic analysis captures much of the unique properties of PCV dynamics, accurate details of the vortex mass are not described, and this mass property arising from core effects plays an essential role in the PCV dynamics. In this paper we give a mean field microscopic derivation of the Turner model using a variational Lagrangian approach. This method provides a systematic way to explore the properties of the PCV mass term. We test the Turner model using numerical simulations of the spin-1 Gross-Pitaevskii equations and find semi-quantitative agreement. In addition we find numerically that the coupling of spin waves to a PCV core may have a substantial effect on the PCV dynamics. Our microscopic treatment offers obvious paths for future investigation that could lead to refinements of the Turner model.

The outline of the paper is as follows. In Sec.~\ref{formsec} we introduce the background formalism for our study of PCVs. In Sec.~\ref{PCVsec} we develop a model of PCV dynamics using a variational Lagrangian approach. In Sec.~\ref{vavsec} we present numerical simulations of a PCV vortex antivortex pair and compare to predictions of the model, which highlights the important role spin waves may play in the PCV dynamics. We conclude in Sec.~\ref{concsec} with a summary and suggestions for future research.

\section{Formalism}\label{formsec}
\subsection{The Spin-1 Gross-Pitaevskii equations} 
The system we consider is a homogeneous quasi-two-dimensional (quasi-2D) spin-1 condensate  
described by the Hamiltonian~\cite{Ho1998a,Ohmi1998a},
\begin{align}\label{spinH}
H=\int d^2\bm{x}\,\left[\bm{\psi}^\dagger \left(-\frac{\hbar^2\nabla^2}{2M}-pf_z+qf_z^2\right)\bm{\psi}+\frac{g_n}{2}n^2+\frac{g_s}{2}\left|\bm{F}\right|^2\right].
\end{align}
Here $\bm{\psi}\equiv (\psi_{1},\psi_0,\psi_{-1})^T$ is a three component spinor describing the condensates in the three spin levels ($m=-1,0,1$) and $p$ and $q$ are, respectively, the linear and quadratic Zeeman shifts arising from the presence of an external field along $z$.  We take the condensate to lie in the $x$-$y$ plane. The term $g_nn^2$ describes the density interaction with coupling constant $g_n$, where  $n\equiv\bm{\psi}^\dagger\bm{\psi}$ is the areal density. 
The term $g_s|\bm{F}|^2$ describes the spin interaction with coupling constant $g_s$, where  $\bm{F}\equiv \bm{\psi}^\dagger\bm{f}\bm{\psi}$  is the areal spin density for the spin-1 matrices $(f_x,f_y,f_z)\equiv\bm{f}$. We take the quantization axis to be along $F_z$, which we choose to be parallel to the position axis $z$. For the system to be mechanically stable we require $g_n>0$. The coupling $g_s$ can be positive or negative resulting in antiferromagnetic or ferromagnetic interactions, respectively.  Here we consider the case of ferromagnetic interactions, i.e.~ $g_s<0$, as realized in $^{87}$Rb  condensates \cite{Chang2004a}. In spinor condensate experiments the quasi-2D regime has been realized by using a trapping potential with tight confinement in one direction (e.g.~see \cite{Sadler2006a}). Our interest is in homogeneous systems where the vortex dynamics will be simpler, noting that recent experiments have realized flat-bottomed optical traps for this purpose \cite{Navon2015a,Chomaz2015a}.

The dynamics of the system can be described by the three coupled Gross-Pitaevskii equations (GPEs),
\begin{align}\label{spinGPEs}
i\hbar\frac{\partial\bm{\psi}}{\partial t}=\left(-\frac{\hbar^2\nabla^2}{2M}-pf_z+qf_z^2+g_nn+g_s\bm{F}\cdot\bm{f}\right)\bm{\psi}.
\end{align}
The linear Zeeman term can be removed by moving to a rotating frame, $\bm{\psi}\rightarrow e^{ipf_zt/\hbar}\bm{\psi}$, so that from hereon we set $p=0$.

The spin interaction energy per particle is on the order of $q_0\equiv 2|g_s|n_0$ where $n_0$ is the mean condensate density. From this energy scale we introduce the spin time,
\begin{align}
t_s\equiv\frac{\hbar}{q_0},
\end{align}
and the spin healing length,
\begin{align}
\xi_s\equiv\frac{\hbar}{\sqrt{Mq_0}}.
\end{align}
The quadratic Zeeman energy per particle we use will also be on the order of $q_0$.

\subsection{Ground states}
\begin{figure}
\includegraphics[trim={13cm 17cm 0cm 26cm},clip=true,width=0.5\textwidth]{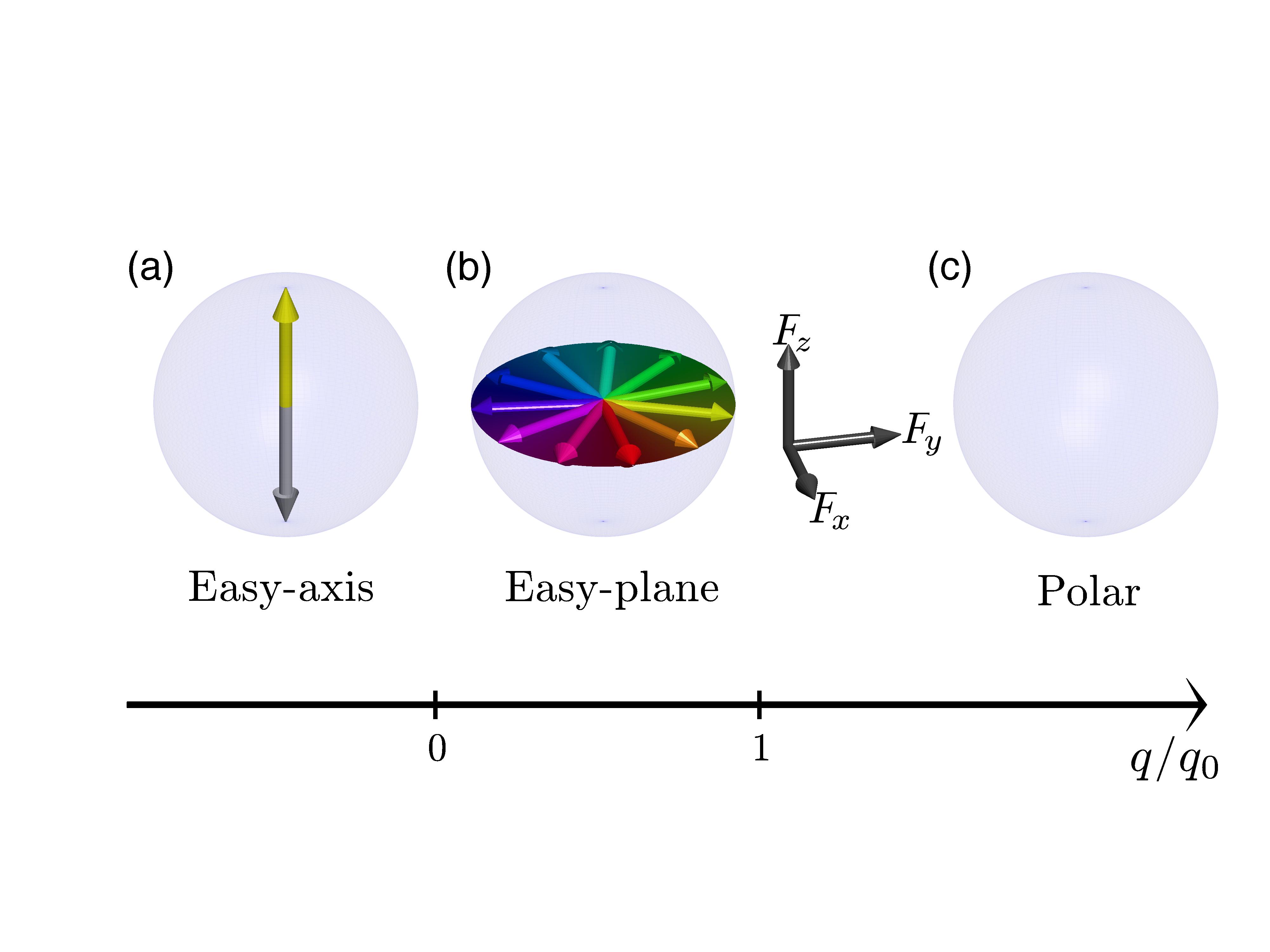}
\caption{\label{phasediag}Ground state magnetic phases for different values of the quadratic Zeeman energy $q$ for an external field along $z$. The direction of the arrows within each sphere signify the ground state magnetization directions. The polar phase has no magnetization. We choose $F_z$ to be parallel to $z$. PCVs can form as defects in the easy plane phase (b).}
\end{figure}

By varying the quadratic Zeeman energy in~\eqref{spinH}, the ground state of the system can be varied between different magnetic states~\cite{Kawaguchi2012R}, see Fig.~\ref{phasediag}. For $0<q<q_0$, the system magnetizes in a plane perpendicular to the applied field, termed the easy plane phase. The ground state wave function of this phase is
\begin{align}\label{psig}
\bm{\psi}_g=\sqrt{n_0}e^{i\theta}\left(\begin{array}{c}\frac{\sin\beta}{\sqrt{2}} e^{-i\phi}\\\cos\beta\\\frac{\sin\beta}{\sqrt{2}}e^{i\phi}\end{array}\right),
\end{align}
where $n_0$ is the (uniform) density, $\cos(2\beta)=q/q_0$, and $\theta$ and $\phi$ are phases arising from the gauge and transverse spin symmetries respectively. The transverse spin symmetry is a result of the invariance of the Hamiltonian~\eqref{spinH} under spin rotations about $F_z$. This wave function gives rise to a spin density
\begin{align}
\bm{F}=n_0\sqrt{1-\left(\frac{q}{q_0}\right)^2}\left(\cos\phi,\sin\phi,0\right),
\end{align}
so that $\phi$ gives the transverse spin angle.

\subsection{Polar-core spin vortices}
Phase winding of the transverse spin angle $\phi$ gives rise to PCVs. This corresponds to a circulation of the superfluid flow of $F_z$ magnetization, since the $F_z$ superfluid current is proportional to $\nabla\phi$~\cite{Yukawa2012}. The core of a PCV is filled by the $m=0$ component. The core is therefore in the polar phase and removes the phase singularity in the transverse magnetization. 

\begin{figure}
\includegraphics[trim={3cm 6cm 3cm 6cm},clip=true,width=0.5\textwidth]{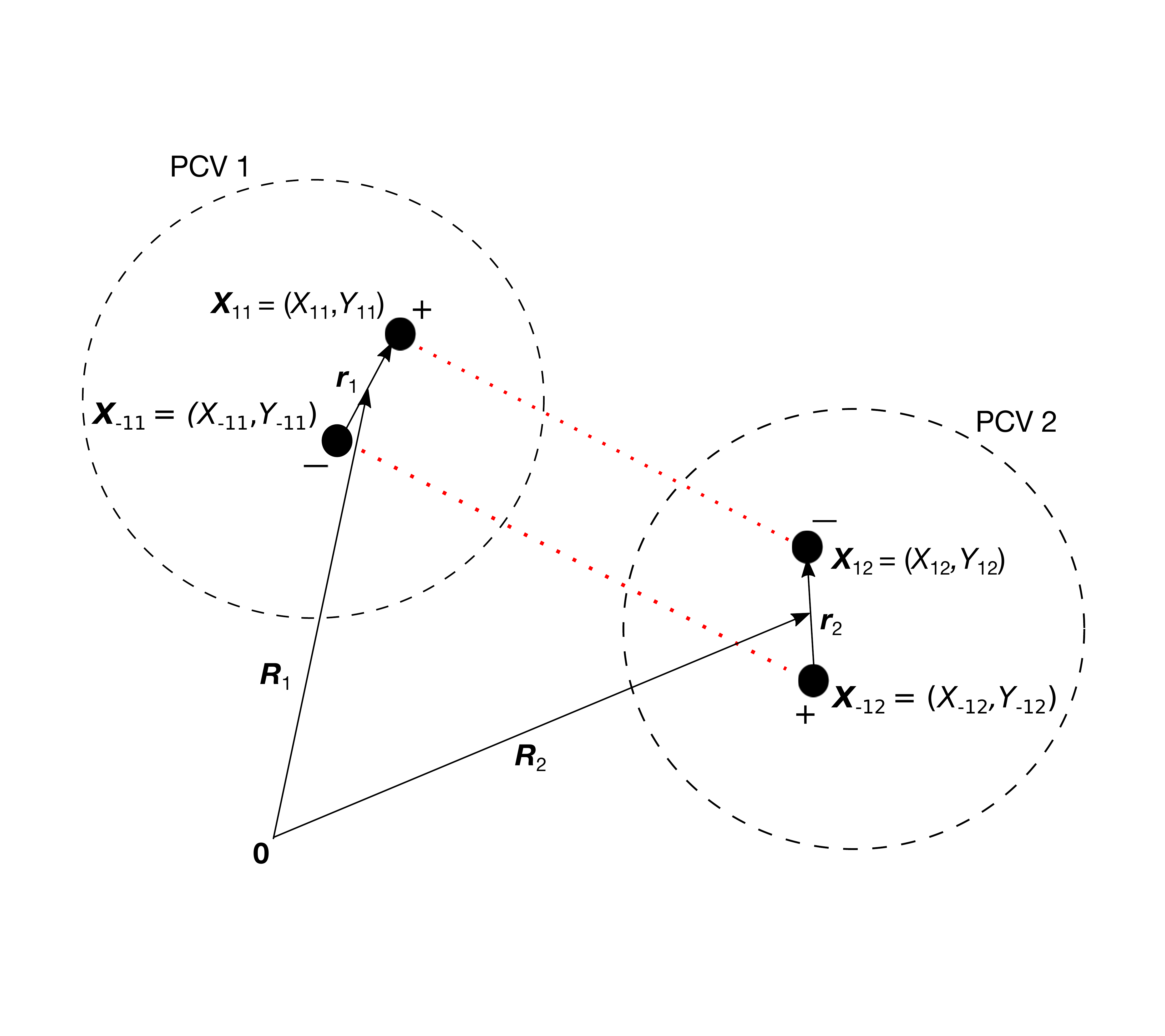}
\caption{\label{PCVschem}Schematic showing two PCVs with coordinates used in this paper. Black filled circles indicate the centre of circulation of the $\psi_{\pm 1}$ vortices within each PCV, with the sign of the vortices displayed by $+$ or $-$. Red dotted lines indicate the interaction between the two $\psi_1$ vortices and the two $\psi_{-1}$ vortices. Around each black dashed circle there will be a phase winding of the transverse spin angle. PCV 1 is a negatively charged vortex ($\kappa_1<0$, phase winding is clockwise) and PCV 2 is a positively charged vortex ($\kappa_2>0$, phase winding is anticlockwise).}
\end{figure}

A PCV is a composite vortex consisting of vortices of opposite charge in the $m=\pm 1$ components, see Eq.~\eqref{psig} and Fig.~\ref{PCVschem}. There is a mass flow of $\psi_1$ around the vortex in the $m=1$ component and an equal but opposite flow of $\psi_{-1}$ around the vortex in the $m=-1$ component. This gives rise to a net flow of $F_z$ magnetization, but no net mass flow.

The state of a single PCV with the centre of circulation of the $\psi_{\pm 1}$ components at the origin can be written as
\begin{align}\label{wansatz1}
\bm{\psi}(\bm{x})=&\sqrt{n_0}\left(\begin{array}{c}g_1(\bm{x})\frac{\sin\beta}{\sqrt{2}} e^{-i\kappa \phi_{1}}\\g_0(\bm{x})\cos\beta \\g_{-1}(\bm{x})\frac{\sin\beta}{\sqrt{2}}e^{i\kappa \phi_{-1}}\end{array}\right),
\end{align}
where $\kappa \phi_{m}$ are phase profiles that give rise to a $2\pi\kappa$ phase winding of the $m$th spin component about the origin and $\kappa\in\mathbb{Z}\setminus\{0\}$ gives the charge of the total PCV. The phase winding of the transverse spin is anticlockwise around a positively charged PCV and clockwise around a negatively charged PCV.

Far from the vortex core the vortex energy density is simply the kinetic energy arising from the phase winding. This will be minimised with a circular phase profile,
\begin{align}\label{circphase}
\phi_{m}=&\operatorname{phase}\left(x+iy\right),
\end{align}
where $\operatorname{phase}(z)$ returns the phase angle of the complex argument $z$. Within the vortex core the phase profile can be modified due to the spin exchange energy, as will be discussed later. To avoid a kinetic energy singularity at the vortex centres, we include (unspecified) real core structures $g_m\ge 0$ in each of the spin components. These core functions should satisfy,
\begin{align}\label{corefunctions}
g_{\pm 1}(\bm{0})=&0,\nonumber\\
g_0(\bm{0})\cos\beta\approx &1,\nonumber\\
g_m(\bm{x})\approx&1\hspace{0.5cm}\text{for $|\bm{x}|>\xi_s$}.
\end{align}
We will not investigate the core structure precisely in this paper. However, we will discuss implications of the core structure. We consider the case of no net $F_z$ magnetization,
\begin{align}\label{Fznet}
\int d^2\bm{x}\,F_z=0.
\end{align}
From symmetry, the core functions of the single PCV~\eqref{wansatz1} will therefore also satisfy $g_1(\bm{x})=g_{-1}(\bm{x})$.

With multiple PCVs, we generalise Eq.~\eqref{wansatz1} to~\cite{fetter1965,bradley2012,lucas2014,eto2011}
\begin{align}\label{wansatz}
\bm{\psi}(\bm{x})=&\sqrt{n_0}\left(\begin{array}{c}\frac{\sin\beta}{\sqrt{2}} \prod_ k g_{1k}(\bm{x})e^{-i\kappa_k\phi_{1k}}\\\cos\beta \prod_kg_{0k}(\bm{x})\\\frac{\sin\beta}{\sqrt{2}}\prod_k g_{-1k}(\bm{x})e^{i\kappa_k\phi_{-1k}}\end{array}\right),
\end{align}
where a subscript $k$ denotes the $k$th PCV. Now the $\kappa_k\phi_{mk}$ are phase profiles that give rise to a $2\pi\kappa_k$ phase winding of the $m$th spin component about the point $\bm{X}_{mk}\equiv (X_{mk},Y_{mk})$. The points $\bm{X}_{mk}$ are centres of circulation, see Fig.~\ref{PCVschem}.

\subsection{Lagrangian formulation}
The fields $i\psi_m(\bm{x},t)$ and $\psi_m^*(\bm{x},t)$ are conjugate fields satisfying Hamilton's equation of motion~\cite{Kawaguchi2012R}
\begin{align}
i\hbar\frac{\partial \psi_m(\bm{x},t)}{\partial t}=\frac{\delta H}{\delta \psi_m^*(\bm{x},t)}.
\end{align}
We obtain the Lagrangian by means of a Legendre transform of the Hamiltonian~\eqref{spinH},
\begin{align}\label{spinL}
L=&i\hbar\sum_m\int d^2\bm{x}\frac{\partial \psi_m(\bm{x},t)}{\partial t}\psi_m^*(\bm{x},t)-H.
\end{align}
We decompose the Lagrangian as follows,
\begin{align}
L=&L_\text{int}+\sum_m L_m,
\end{align}
where
\begin{align}\label{Lm}
L_m\equiv \int d^2\bm{x}\,\psi_m^*\left[i\hbar\frac{\partial}{\partial t}+\frac{\hbar^2\nabla^2}{2M}\right]\psi_m,
\end{align}
and
\begin{align}\label{Lcomp}
L_\text{int}\equiv -\int d^2\bm{x}\,\left[\frac{g_n}{2}n^2+\frac{g_s}{2}\left|\bm{F}\right|^2+q\left(|\psi_1|^2+|\psi_{-1}|^2\right)\right],
\end{align}

\section{A model of vortex dynamics}\label{PCVsec}
\subsection{The variational Lagrangian}
The dynamics of PCVs governed by the GPEs are complicated by the possibility of dynamic core effects and coupling between PCVs and other excitations such as spin waves. These effects are ignored in Turner's model, and we will do the same here. We use a variational approach using the wavefunction ansatz~\eqref{wansatz} with the vortex positions $\bm{X}_{mk}(t)=(X_{m k}(t),Y_{m k}(t))$ as variational parameters. We assume static core structures
\begin{align}
g_{mk}(\bm{x})=g_m\left(\bm{x}-\bm{X}_{mk}(t)\right),
\end{align}
and static phase profiles. Assuming static core structures and phase profiles neglects possibly interesting core dynamics. For example, we will find that coupling to spin waves may have an important effect on the PCV dynamics. However, essential parts of the PCV dynamics are still captured within the approximations made here.

We substitute~\eqref{wansatz} into the Lagrangian~\eqref{spinL} and minimise the action $S=\int dt\,L$ with respect to the variational parameters $(X_{mk},Y_{mk})$. This is equivalent to assuming that the variational parameters obey Lagrange's equations of motion.

The terms $L_m$~\eqref{Lm} will be dominated by phase variation outside the vortex cores. We therefore ignore the core structures in the evaluation of these terms and assume circular phase profiles~\eqref{circphase}. The $L_m$ terms are then identical to the Lagrangian for three non-interacting scalar condensates and, after appropriate regularization, become (see e.g.~\cite{lucas2014})
\begin{align}\label{Ls}
L_m=&-\frac{\alpha}{2}\frac{M}{\hbar}\sum_km\kappa_k\left(\dot{X}_{mk}Y_{mk}-\dot{Y}_{mk}X_{mk}\right)\nonumber\\
&+\frac{\alpha}{2}\sum_{j,k>j} \kappa_j\kappa_k\ln\left|\frac{\bm{X}_{m k}-\bm{X}_{m j}}{l}\right|^2,
\end{align}
where
\begin{align}
\alpha\equiv\frac{\pi\hbar^2 n_0\sin^2\beta}{2M},
\end{align}
and $l$ is the system size.

The term $L_\text{int}$~\eqref{Lcomp} can be written as
\begin{align}\label{Lintv}
L_\text{int}=L_\text{D}+L_\text{SE}.
\end{align}
The density term $L_\text{D}$ is
\begin{align}\label{Lddeq}
L_\text{D}=&(g_s-g_n)\int d^2\bm{x}\,|\psi_1|^2|\psi_{-1}|^2\nonumber\\
&-(g_s+g_n)\int d^2\bm{x}\,|\psi_0|^2\left(|\psi_1|^2+|\psi_{-1}|^2\right).
\end{align}
The terms in $L_\text{D}$ depend only on the core structures $g_m$ and will be constant for $g_m(\bm{x})=1$. We have ignored $|\psi_m|^2$ and $|\psi_m|^4$ terms in~\eqref{Lddeq}, as these only contribute a constant term for static core structures $g_m$ and nonoverlapping PCV cores. The remaining \emph{spin exchange} term is
\begin{align}
L_\text{SE}=&-2g_s\operatorname{Re}\int d^2\bm{x}\,\psi_1^*\psi_{-1}^*\psi_0^2,\nonumber\\
=&-g_s n_0^2\sin^2\beta\cos^2\beta\nonumber\\
&\times\int d^2\bm{x}\,\prod_kg_{1k}g_{-1k}g_{0k}^2\cos\left(\kappa_k\phi_{1k}-\kappa_k\phi_{-1k}\right).
\end{align}
This spin exchange interaction depends not only on the core structures $g_m$ but also on the phase profiles $\phi_{\pm 1k}$. This term will therefore affect the dynamics even with negligible core structures, i.e. with $g_m(\bm{x})=1$. The spin exchange interaction allows for two $m=0$ atoms to scatter into $m=\pm 1$ atoms, and vice versa. This interaction is not present in a multi-component condensate of different atoms.

We are now in a position where we can write down equations of motion for the variational parameters $(X_{m k},Y_{m k})$. We do this implicitly by introducing ``centre of mass'' and relative coordinates for each PCV (see Fig.~\ref{PCVschem}),
\begin{align}\label{cmrc}
\bm{R}_k=&\frac{\bm{X}_{1 k}+\bm{X}_{-1 k}}{2}, \\
\bm{r}_k=&\bm{X}_{1k}-\bm{X}_{-1k}.
\end{align}
We assume dynamics where the spacing between vortex $k$ and the remaining PCVs is much larger than the relative coordinate $r_k$, i.e. $r_k\ll |\bm{R}_k-\bm{R}_j|$. We can therefore set $|\bm{X}_{1k}-\bm{X}_{-1j}|\approx |\bm{X}_{-1k}-\bm{X}_{-1j}|\approx |\bm{R}_k-\bm{R}_j|$ and talk about a single PCV being at position $\bm{R}_k$.

This gives the full Lagrangian
\begin{align}\label{Lfull}
L=&\frac{\alpha}{2}\Bigg[-\frac{M}{\hbar}\sum_k \kappa_k\left(\dot{X}_ky_k-\dot{y}_kX_k-\dot{Y}_kx_k+\dot{x_k}Y_k\right)\nonumber\\
&+2\sum_{j,k>j} \kappa_j\kappa_k\ln\left|\frac{\bm{R}_k-\bm{R}_j}{l}\right|^2+f_\text{D}+f_\text{SE}\Bigg],
\end{align}
where
\begin{align}
f_\text{D}\equiv\frac{2}{\alpha}L_\text{D},
\end{align}
and
\begin{align}
f_\text{SE}\equiv\frac{2\cos^2\beta}{\pi\xi_s^2}\int d^2\bm{x}\,\prod_kg_{1k}g_{-1k}g_{0k}^2\cos\left(\kappa_k\phi_{1k}-\kappa_k\phi_{-1k}\right).
\end{align}
Lagrange's equations of motion are
\begin{align}
\frac{\partial L}{\partial \chi_k}=\frac{d}{dt}\frac{\partial L}{\partial \dot{\chi}_k},
\end{align}
for the variational parameters $\chi_k\in\{X_k,Y_k,x_k,y_k\}$. The momentum conjugate to a spatial coordinate $\chi$ is $p_\chi\equiv\partial L/\partial\dot{\chi}$. The Lagrangian~\eqref{Lfull} therefore gives,
\begin{subequations}\label{canrel}
\begin{align}
p_{X_k}=&-\frac{M\alpha}{2\hbar}\kappa_k y_k, \\
p_{Y_k}=&\frac{M\alpha}{2\hbar}\kappa_k x_k, \\
p_{x_k}=&-\frac{M\alpha}{2\hbar}\kappa_k Y_k, \\
p_{y_k}=&\frac{M\alpha}{2\hbar}\kappa_k X_k.
\end{align}
\end{subequations}

The conjugate variable relations~\eqref{canrel} show that a centre of mass coordinate $\bm{R}_k$ is conjugate to the rotated relative coordinate $\bm{r}_k\times\hat{\bm{z}}$. This differs from the scalar vortex case, where the $X$ and $Y$ positions of a single vortex are conjugate pairs~\cite{aref2007}. The four Lagrange's equations of motion that describe the PCV dynamics are\begin{subequations}\label{Lageq}
\begin{align}
\dot{X}_k=&\frac{\hbar}{2M \kappa_k}\frac{\partial (f_\text{D}+f_\text{SE})}{\partial y_k},\\
\dot{Y}_k=&-\frac{\hbar}{2M\kappa_k}\frac{\partial (f_\text{D}+f_\text{SE})}{\partial x_k}, \\
\dot{y}_k=&-\frac{\hbar}{M}\left[2\sum_{j\ne k}\kappa_j\frac{X_k-X_j}{|\bm{R}_k-\bm{R}_j|^2}+\frac{1}{2\kappa_k}\frac{\partial (f_\text{D}+f_\text{SE})}{\partial X_k}\right], \\
\dot{x}_k=&\frac{\hbar}{M}\left[2\sum_{j\ne k}\kappa_j\frac{Y_k-Y_j}{|\bm{R}_k-\bm{R}_j|^2}+\frac{1}{2\kappa_k}\frac{\partial (f_\text{D}+f_\text{SE})}{\partial Y_k}\right].
\end{align}\end{subequations}

\subsection{Conservation laws}
Lagrange's equations of motion~\eqref{Lageq} are invariant under translations in space, translations in time and rotations. Invoking Noether's theorem, these invariances give rise to the following conserved quantities.
Invariance under translations in time gives rise to conservation of the Hamiltonian
\begin{align}\label{Htot}
H=&-\frac{\alpha}{2}\left[2\sum_{j,k>j} \kappa_j\kappa_k\ln\left|\frac{\bm{R}_k-\bm{R}_j}{l}\right|^2+f_\text{D}+f_\text{SE}\right].
\end{align}
Invariance under spatial translations gives rise to conservation of the total linear momentum
\begin{align}\label{Ptot}
\bm{P}=&\sum_k\left(p_{X_k},p_{Y_k}\right)=\frac{M\alpha}{2\hbar}\sum_k\kappa_k \hat{\bm{z}}\times\bm{r}_k.
\end{align}
Invariance under rotations gives rise to conservation of the total angular momentum
\begin{align}\label{Ltot}
\bm{L}=&\sum_k\left[\left(X_k,Y_k,0\right)\times\left(p_{X_k},p_{Y_k},0\right)+\left(x_k,y_k,0\right)\times\left(p_{x_k},p_{y_k},0\right)\right],\nonumber\\
=&\frac{M\alpha}{\hbar}\sum_k\kappa_k\left(\bm{r}_k\cdot\bm{R}_k\right)\hat{\bm{z}}.
\end{align}

\subsection{Reduction to Turner's model of PCV dynamics}

In scalar vortex dynamics, the vortex cores can be neglected to obtain a point vortex model~\cite{lucas2014}. In contrast, PCV dynamics depend crucially on the internal vortex coordinates $\bm{r}_k$. Although the vortex cores of single spin levels are approximated as static, the total core structure of the PCV can still exhibit dynamics. This arises from the term $f_\text{D}+f_\text{SE}$ in the Lagrangian~\eqref{Lfull}, which makes dealing with the Lagrangian~\eqref{Lfull} numerically complicated. Furthermore, this term depends crucially on the details of the core structures $g_m$. To make progress, we follow~\cite{turner2009} and assume a simple form for this term (ignoring a constant shift),
\begin{align}\label{fsimp}
f_\text{D}+f_\text{SE}=-\frac{a}{2\xi_s^2}\sum_k \kappa_k^2 r_k^2,
\end{align}
where $a$ is an empirical parameter that we will find to be on the order of 0.1. In this approximation we are assuming that the interaction energy of a single PCV core depends only on the coordinate $r_k$, and not on the details of any other PCVs.  (This assumption also allows us to neglect the direction of $\bm{r}_k$.) Furthermore, we are assuming the interaction energy does not deviate much from the energy minimum at $r_k=0$, so that we consider terms up to second order in $r_k$ only. We expect the interaction to be an analytic, even function of $\bm{r}_k$ so that there is no linear term in the expansion. We will term the interaction energy the ``stretch energy'', as it arises from the energy cost of separating the $\psi_1$ and $\psi_{-1}$ components of a PCV, which stretches the PCV core.

If we suppose that the stretch energy can be considered quadratic for $r_k\lesssim\xi_s$, then the quadratic approximation will hold for stretch energies $\lesssim \hbar^2 n_0/M$, see Eq.~\eqref{Htot}. The energy required to stretch the PCV core comes from changes in the logarithmic interaction term in~\eqref{Htot} prior to vortex annihilation. Therefore the stretch energy will remain on the order of $\hbar^2 n_0/M$ as long as $\ln(D_i/\xi_s)$ is not much larger than 1, for initial vortex separations $D_i$ and zero initial stretching. It is plausible that the parameter $a$ may also have an additional dependence on the magnitude of $\kappa_k$. We do not consider this possibility here, and note that if all vortices have the same magnitude of charge then this dependence will not impact the model apart from a fixed change of $a$. We expect that the PCV core structure and phase profile close to the core will depend on the quadratic Zeeman energy $q$, and therefore $a$ will depend on $q$ also.

With the Lagrangian simplified by Eq.~\eqref{fsimp}, we can evaluate Lagrange's equations of motion~\eqref{Lageq}. The two coupled first order equations of motion for $\bm{R}_k$ and $\bm{r}_k$ can be decoupled by taking a second time derivative of $\bm{R}_k$. We then obtain a single equation of motion
\begin{align}\label{Rcol1}
\frac{d^2\bm{R}_k}{dt^2}=&\frac{\hbar^2 a}{M^2\xi_s^2}\sum_{j\ne k}\kappa_k\kappa_j \frac{\bm{R}_k-\bm{R}_j}{|\bm{R}_k-\bm{R}_j|^2},\\
\bm{r}_k=&\frac{2M\xi_s^2}{\hbar a\kappa_k}\frac{d\bm{R}_k}{dt}\times\hat{\bm{z}}.\label{rkeq1}
\end{align}
We can write this in a form analogous to a single particle classical mechanics model by noting that $r_k\propto \left|d\bm{R}_k/dt\right|$ so that the stretch energy (ignoring a constant shift) can be written as,
\begin{align}\label{pvmK}
H_\text{s}\equiv&-\frac{\alpha}{2}\left(f_D+f_\text{SE}\right),\nonumber\\
=&\frac{\alpha a}{4\xi_s^2}\sum_k \kappa_k^2 r_k^2,\nonumber\\
\equiv&\frac{1}{2}m_v\sum_k\left|\frac{d\bm{R}_k}{dt}\right|^2,
\end{align}
with the vortex mass $m_v$ defined as
\begin{align}\label{mdef}
m_v\equiv\frac{\pi Mn_0\xi_s^2\sin^2\beta}{a}.
\end{align}
Equations~\eqref{Rcol1},~\eqref{rkeq1} can then be written as
\begin{align}\label{Rcol}
m_v\frac{d^2\bm{R}_k}{dt^2}&=2\alpha\sum_{j\ne k} \kappa_k \kappa_j\frac{\bm{R}_k-\bm{R}_j}{|\bm{R}_k-\bm{R}_j|^2},\\
\bm{r}_k&=\frac{\hbar m_v}{M \alpha\kappa_k}\frac{d\bm{R}_k}{dt}\times\hat{\bm{z}}. \label{rkeq}
\end{align}
Equation~\eqref{Rcol} resembles the dynamics of classical charged particles of mass $m_v$ moving under the two dimensional Coulomb interaction
\begin{align}\label{pvmU}
U=-\alpha\sum_{j,k>j} \kappa_k \kappa_j\ln\left|\frac{\bm{R}_k-\bm{R}_j}{l}\right|^2.
\end{align}
The Coulomb interaction is the kinetic energy of the condensate arising from the phase winding of the vortices. The total energy $H_\text{s}+U$ will be conserved under the dynamics of Eqs.~\eqref{Rcol},~\eqref{rkeq}. This is the model proposed in~\cite{turner2009}\footnote{We note that our ``Coulomb constant'' is larger than that in~\cite{turner2009} by a factor of $\pi$. However, an expression for the PCV mass is not given in~\cite{turner2009}, and so if our PCV mass~\eqref{mdef} is also larger by a factor of $\pi$ then the relevant quantity $\alpha/m_v$ would be the same as that in~\cite{turner2009}.}. The total linear momentum~\eqref{Ptot} can be written as
\begin{align}
\bm{P}=\frac{1}{2}\sum_k m_v\frac{d\bm{R}_k}{dt}.
\end{align}
The total angular  total angular momentum~\eqref{Ltot} can be written as
\begin{align}
\bm{L}=\sum_k \bm{R}_k\times m_v\frac{d\bm{R}_k}{dt}.
\end{align}

\section{\label{vavsec}Numerical results for a polar core vortex antivortex pair}
\subsection{\label{vavsec1}GPE dynamics of a vortex antivortex pair}
\begin{figure*}
\includegraphics[width=\textwidth]{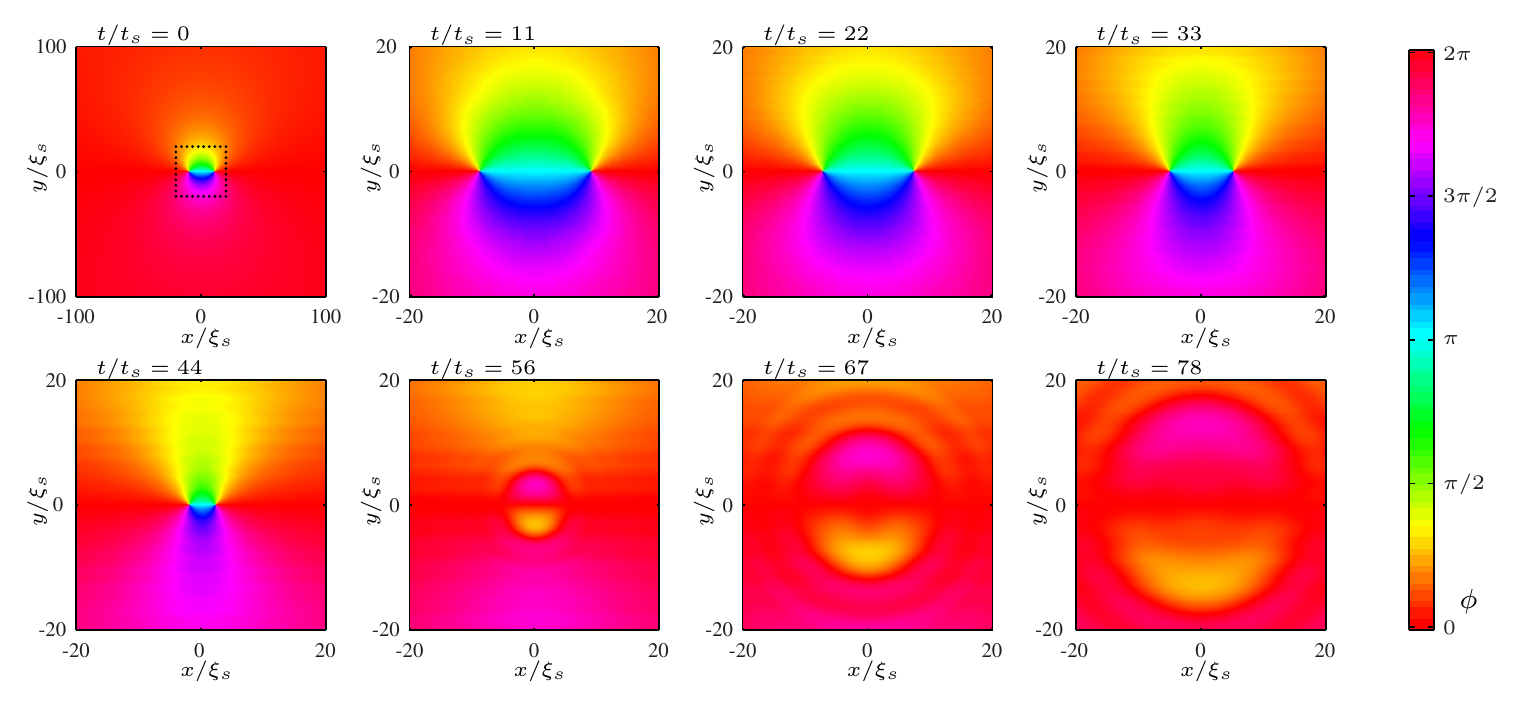}
\caption{\label{vIms}Evolution of transverse spin angle $\phi$ with time. First frame shows the full system while subsequent frames show the area within the dotted square marked in the first frame. Clearly visible are the two oppositely charged PCVs attracting, colliding and annihilating after a time $t\approx 50t_s$. The average speed of the vortex is on the order of the characteristic spin wave speed $\xi_s/t_s$, see Eq.~\eqref{tvdef}. After annihilation a spin pulse propagates away from the vortex collision with a speed on the order of the characteristic spin wave speed.}
\end{figure*}

We consider the evolution of the following initial state,
\begin{align}\label{psiint}
\bm{\psi}=\sqrt{n_0}\left(\begin{array}{c}\frac{\sin\beta}{\sqrt{2}} e^{-i(\phi_2-\phi_1)}\\\cos\beta\\\frac{\sin\beta}{\sqrt{2}}e^{i(\phi_2-\phi_1)}\end{array}\right)
\end{align}
where
\begin{align}
\phi_1(x,y)=&\operatorname{phase}\left(x+\frac{D}{2}+iy\right),\nonumber\\
\phi_2(x,y)=&\operatorname{phase}\left(x-\frac{D}{2}+iy\right),
\end{align}
gives rise to two oppositely charged PCVs with centres of circulation at positions $\bm{R}_1=(-D/2,0)$ (charge $\kappa_1=-1$) and $\bm{R}_2=(D/2,0)$ (charge $\kappa_2=1$), see Fig.~\ref{vIms}. Although higher charged vortices are possible they will be energetically unstable, and will decay into singly charge vortices that preserve topological charge. We do not include an initial core structure in~\eqref{psiint}. However, an appropriate core structure will form from the GPE evolution through the interaction with other excitations, producing a small number of excitations on top of the vortex state. The vortices are initially unstretched, i.e. $\bm{r}_1=\bm{r}_2=\bm{0}$. Note that the initial state~\eqref{psiint} has no net $F_z$ magnetization, Eq.~\eqref{Fznet}.

We solve Eq~\eqref{spinGPEs} with the initial condition~\eqref{psiint} with periodic boundary conditions. We use an adaptive step Runge-Kutta method that uses Fast Fourier transforms to evaluate the kinetic energy operators with spectral accuracy. We choose $q=0.3q_0$, $n_0=10^4/\xi_s^2$ and $g_n/|g_s|=10$. We use a $4096\times 4096$ grid with side lengths $l=200\xi_s$ and initial vortex separation $D=20\xi_s$. With this initial condition there is a discontinuity in the phase profile along the horizontal boundaries. This gives rise to a non-zero kinetic energy at these boundaries, which will manifest itself as spin waves and interact with the vortices. To mitigate this effect we choose $D\ll l$, since far from a dipole the phase profile decays quadratically with distance. Periodic boundary conditions also introduce image charges and prevent simulating systems with a net phase winding around the boundary, for example two same charged PCVs.

\begin{figure}
\includegraphics[width=0.5\textwidth]{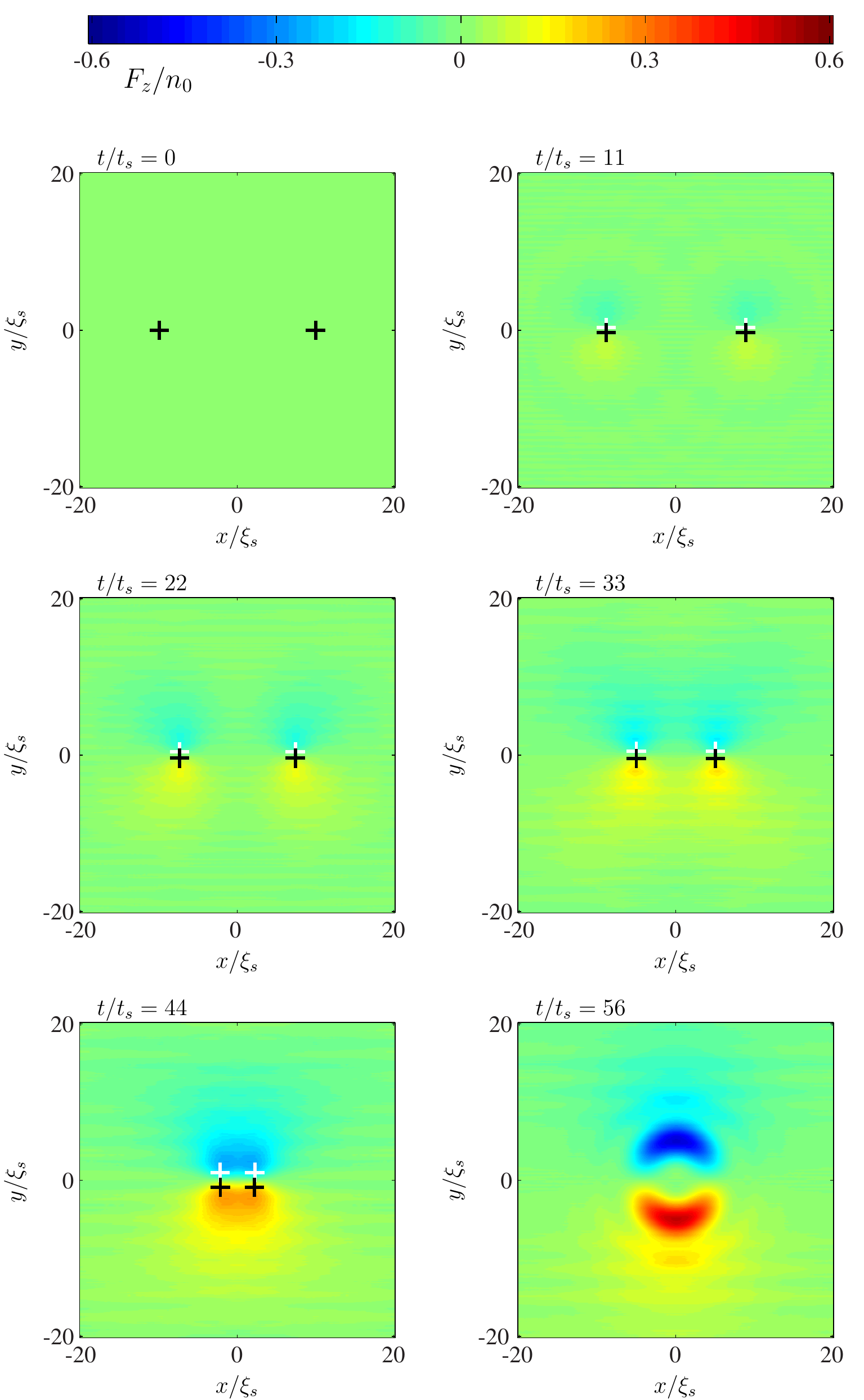}
\caption{\label{stretchIm}Plot of $F_z$ magnetization, Eq.~\eqref{Fzeq}, for various times during the PCV collision. For frames with $t<50 t_s$, the peak and dip in $F_z$ arise from the separation of the cores of the $\psi_{\pm 1}$ vortices. This indicates that each PCV stretches in a direction perpendicular to its centre of mass motion as the PCVs attract. The centre of circulation for vortices in the $\psi_1$ ($\psi_{-1}$) components, i.e. $\bm{X}_{1 k}$ ($\bm{X}_{-1k}$), are marked by white (black) crosses. The direction of stretching is consistent with Eq.~\eqref{rkeq}.}
\end{figure}

Figure~\ref{vIms} shows the evolution of the angle of the transverse spin. As predicted by Eq.~\eqref{Rcol}, the oppositely charged PCVs attract\footnote{Note that the PCV dynamics is very different from scalar vortex dynamics. In the absence of sound waves, two equal but oppositely charged scalar vortices will move parallel to one another rather than attracting~\cite{fetter1965,jones1982}.}. The vortices annihilate at time $t\approx 50t_s$ and their energy is liberated as a propagating spin pulse (frames with $t>50t_s$). The pulse travels at a speed on the order of the characteristic spin wave speed $\xi_s/t_s$~\cite{symes2014b}. Associated with this collision is a stretching of the PCVs perpendicular to their centre of mass velocity, as predicted by Eq.~\eqref{rkeq}. We show this stretching in Fig.~\ref{stretchIm} by plotting the $F_z$ magnetization
\begin{align}\label{Fzeq}
F_z=|\psi_1|^2-|\psi_{-1}|^2.
\end{align}
A separation of the positive $F_z$ peak from the negative $F_z$ dip within each PCV core arises from the separation of the cores of the $\psi_{\pm 1}$ vortices. Included in this figure are white (black) crosses that mark the centre of circulation for vortices in the $\psi_1$ ($\psi_{-1}$) components i.e. the $\bm{X}_{1 k}$ ($\bm{X}_{-1k}$) points. The direction of stretching is consistent with Eq.~\eqref{rkeq}.

The initial state~\eqref{psiint} gives rise to PCVs with centre of mass positions $\bm{R}_1=(-D/2,0)$ (charge $\kappa_1=-1$) and $\bm{R}_2=(D/2,0)$ (charge $\kappa_2=1$), both with zero initial stretching, $\bm{r}_1=\bm{r}_2=\bm{0}$. With this initial condition, Eqs.~\eqref{Rcol},~\eqref{rkeq} can be solved analytically to obtain\footnote{To see where this solution comes from, note again that the equation of motion for $D$ resembles that of a classical particle moving in a conservative potential. The equation of energy conservation can be rearranged to obtain an equation of motion for $\dot{D}$, which can be solved by integration to obtain $t$ as a function of $D$~\cite{taylor2005}. Inverting gives $D$ as a function of $t$.}
 
\begin{align}
D(t)=&D(0)\exp\left[-\left(\operatorname{erf}^{-1} \left(\frac{t}{t_\text{coll}}\right)\right)^2\right],\label{Dan}\\
r(t)=&\frac{\hbar}{M}\sqrt{\frac{2m_v}{\alpha}}\operatorname{erf}^{-1}\left(\frac{t}{t_\text{coll}}\right).\label{Dan2}
\end{align}
Here
\begin{align}\label{tvdef}
t_\text{coll}\equiv &D(0)\sqrt{\frac{\pi m_v}{8\alpha}}\sim\frac{t_s}{\xi_s}D(0),
\end{align}
gives the time the vortices take to collide. Note that this solution predicts that the average speed of the vortex should be independent of the initial separation and should be on the order of the characteristic spin wave speed $\xi_s/t_s$. The image charges that are introduced in the numerics from the periodic boundary conditions could be accounted for when solving Eqs.~\eqref{Rcol} and~\eqref{rkeq}~\cite{weiss1991}. For the small ratios of $D/l$ in Fig.~\ref{vIms}, the image charges will have only a small effect on the dynamics and so we ignore them here.

\begin{figure}
\includegraphics[width=0.5\textwidth]{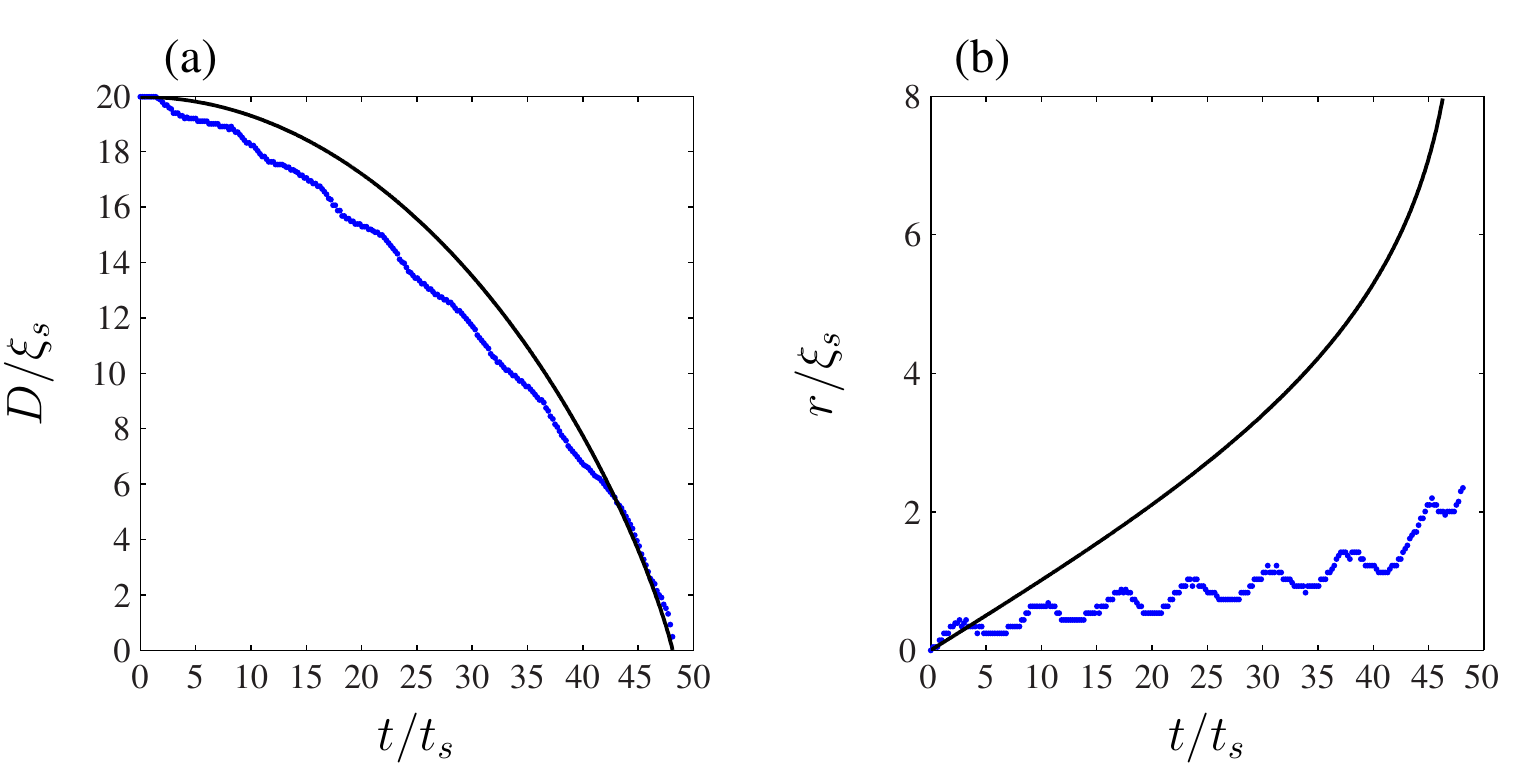}
\caption{\label{vdata}(a) PCV separation versus time. Blue dots are numerical data, solid black line is the analytic prediction~\eqref{Dan} with $m_v$ as a fitting parameter. (b) Separation of the $\psi_{\pm 1}$ vortex centres versus time for a single PCV. Blue dots are numerical data, solid black line is the analytic prediction~\eqref{Dan2} using $m_v$ obtained from the fit in (a). For a short time $t\lesssim 3t_s$ during the initial evolution, the numerical result in (b) lies slightly above the analytic prediction. After this, the numerical result lies below the analytic prediction. Also visible in the numerical results in (a) and (b) is an oscillation at a frequency on the order of the characteristic spin wave frequency, $\omega_s\equiv 1/t_s$, which is not predicted by the analytic model~\eqref{Dan}, \eqref{Dan2}. We suspect the oscillation arises from coupling to spin waves. As a PCV stretches, interactions within the core may couple the core to other spin excitations so that energy is lost in the form of spin waves, thus reducing the stretching below the analytic prediction.}
\end{figure}

Fig.~\ref{vdata}(a) shows the numerical result for the PCV separation $D$ against time. Also included in Fig.~\ref{vdata}(a) is the analytic solution~\eqref{Dan} with the vortex mass as a fitting parameter. From this fit we obtain that $a\approx 0.135$ so that the vortex mass comes out as $m_v\sim 10 M n_0^2\xi_s^2$, which is roughly an order of magnitude larger than the physical mass of the vortex core. We have carried out similar simulations for interaction ratios $g_n/|g_s|=2.5$ and $g_n/|g_s|=40$, which also gave $a\approx 0.135$. We find that fluctuations of total density outside the PCV cores are essentially zero and within the PCV cores are small, since the relevant energy scale of our system, $|g_s|n_0$, is much less than the energy scale of density fluctuations, $g_n n_0$. In experiments with $^{87}$Rb, $g_n/|g_s|\sim 100$~\cite{Kawaguchi2012R} so that density fluctuations will be even smaller.

Fig.~\ref{vdata}(b) shows the numerical result for the vortex stretching against time, along with the analytic prediction~\eqref{Dan2} using the fitted mass from~(a). For a short time $t\lesssim 3t_s$ during the initial evolution, the numerical result lies slightly above the analytic prediction. After this, the numerical result lies below the analytic prediction. Also visible in the numerical results (a) and (b) is an oscillation with a period on the order of $2\pi t_s$. We suspect that these oscillations arise from coupling to spin waves, a possibility that was also suggested in~\cite{turner2009}. As a PCV stretches, it is possible that interaction effects inside the PCV core couple the core to other spin excitations, allowing the stretch energy to leave the core as spin waves. The analytic prediction in Fig.~\ref{vdata}(b) would then be expected to be larger than the numerical result since energy lost to other excitations is not accounted for in the analytic model~\eqref{Rcol}, \eqref{rkeq} but is in the GPE dynamics\footnote{Periodic boundary conditions generate a discontinuity in the initial phase profile along the horizontal boundaries. As the system evolves, this discontinuity will smooth out and generate additional excitations during the vortex evolution, which may also influence the PCV dynamics. Excitations will also be produced in the initial GPE evolution as the PCV cores form.}. The possible spin waves include transverse and axial spin excitations. Additional core dynamics may also be present that do not result in the emission of spin waves, for example spin oscillations that change the core structure with time while preserving the core energy. Note that the time scale of vortex dynamics is comparable to the time scale of the spin interaction energy, so that PCV motion will likely not be adiabatic compared to core dynamics arising from the spin interaction.

The analytic results in Figs.~\ref{vdata}(a) and (b) capture the main qualitative behaviours of PCVs, namely that PCVs of opposite charge attract and stretch perpendicular to their centre of mass motion as they accelerate toward each other. Therefore the analytic model captures the drastic qualitative difference between the dynamics of PCVs and scalar vortices. The analytic model overestimates the PCV stretching but does give a reasonable estimate for the time scale of the vortex annihilation. Finding accurate vortex core ground states (e.g.~see \cite{Isoshima2001a}) could be used to determine the interaction term~\eqref{Lintv} more accurately. The effect of spin waves has been neglected in the analytic model presented here. Including the effect of spin waves exactly would likely be difficult. However it may be possible to approximate this effect, or include the effect phenomenologically.

\subsection{The spin exchange interaction energy and confinement}

\begin{figure}
\includegraphics[width=0.5\textwidth]{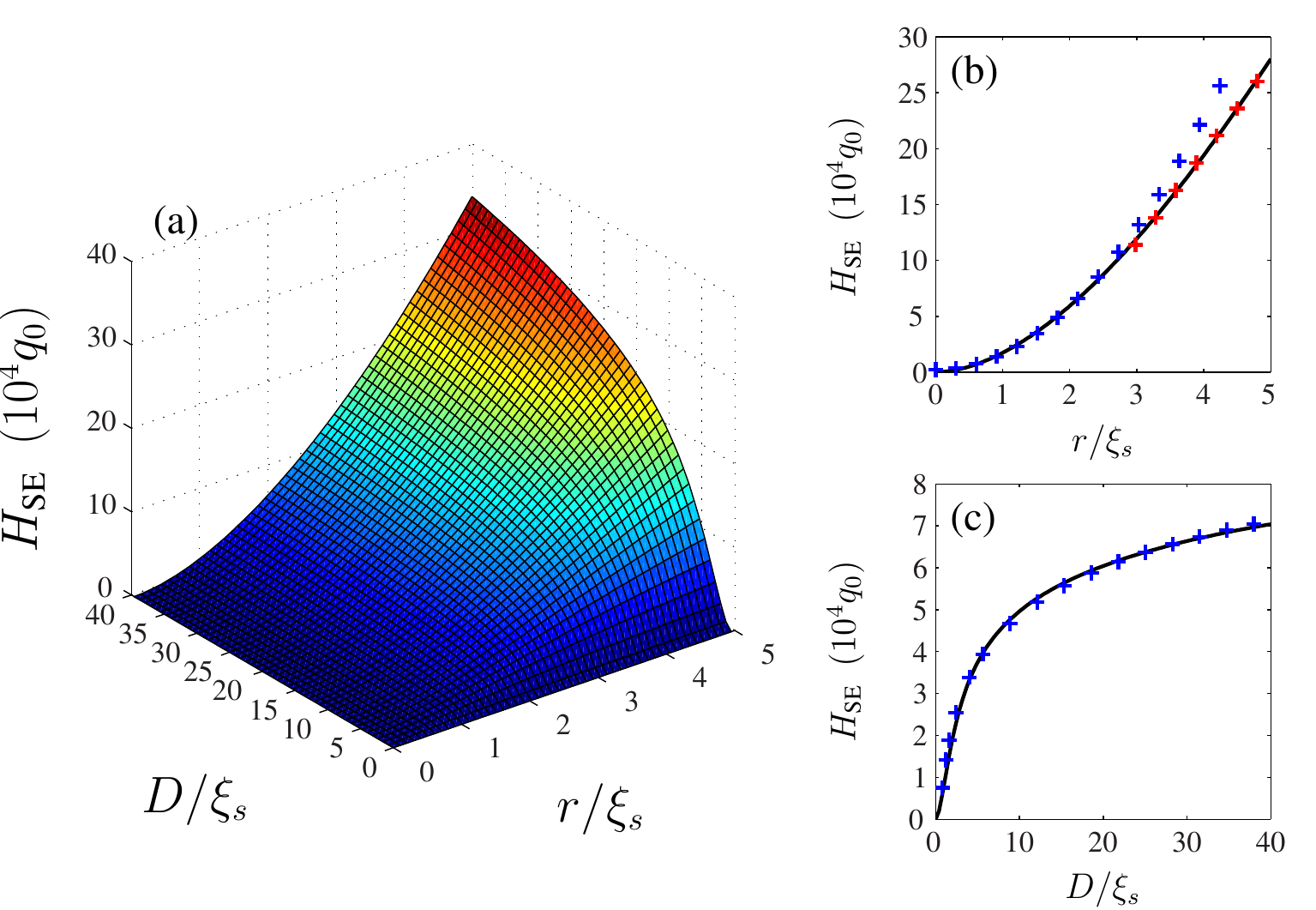}
\caption{\label{stretchE}(a) The spin exchange energy, Eq.~\eqref{HSEapprox}, as a function of the PCV separation $D$ and the PCV stretching $r$. (b) Solid black line shows data from (a) for $D=20\xi_s$, as a function of $r$. For $r\lesssim 3\xi_s$ the energy increases quadratically with $r$. A quadratic fit to the data at small $r$ is shown by blue crosses. For $r>3\xi_s$ the energy increases linearly with $r$. A linear fit to the data at large $r$ is shown by red crosses. (c) Solid black line shows data from (a) for $r=2\xi_s$, as a function of $D$. The energy increases logarithmically with $D$. A logarithmic fit to the data is shown by blue crosses.}
\end{figure}

The interaction term~\eqref{Lintv} depends on both the internal phase profile of a PCV and the core structure. For $|\bm{x}|\gg \xi_s$, we have $g_m(\bm{x})\approx 1$ [see Eq.~\eqref{corefunctions}] and the vortex phase profile will be close to circular. In this regime only the spin-exchange part of the interaction term~\eqref{Lintv} will affect the dynamics. For the case of two singly charged PCVs of opposite charge, the energy corresponding to this term takes the form
\begin{align}\label{HSEapprox}
H_\text{SE}\approx \frac{\alpha\cos^2\beta}{\pi\xi_s^2}\int d^2\bm{x}\,\left(1-\prod_{k=1,2}\cos\left(\phi_{1k}-\phi_{-1k}\right)\right),
\end{align}
for circular phase profiles $\phi_{\pm 1k}$, Eq.~\eqref{circphase}. We take the zero point energy to be the energy of two unstretched PCVs. Fig.~\ref{stretchE}(a) shows a plot of the spin-exchange energy~\eqref{HSEapprox} versus the vortex separation, $D$, and vortex stretching, $r$, obtained by numerically integrating Eq.~\eqref{HSEapprox}. For each evaluation of the energy, we fix $X_{\pm 1 1}=-X_{\pm 1 2}=D/2$ and $Y_{\pm 1 k}=\pm r/2$. Note the stretching is orthogonal to the vortex separation. Fig.~\ref{stretchE}(b) shows the dependence of the spin exchange energy on the vortex stretching for $D=20\xi_s$. These results depend only very weakly on total system size as long as the system size is large, since the four component vortices present will behave like a quadrupole with a rapidly decaying energy density for distances greater than $D$. We find that for small stretching, $r<3\xi_s$, the energy is quadratic in $r$. For larger stretching the energy becomes linear in $r$. Fig.~\ref{stretchE}(c) shows the dependence of the spin exchange energy on the vortex separation $D$ for $r=2\xi_s$. The dependence follows a logarithmic form very closely over the range of $D$ values considered. This logarithmic dependence on $D$ can be argued heuristically by assuming that the spin exchange energy density scales as
\begin{align}
\mathcal{H}(x,y)\sim\Bigg\{\begin{array}{cc}\frac{r^2}{x^2+y^2},&x^2+y^2<D^2,\\0,&x^2+y^2>D^2.\end{array}\nonumber
\end{align}
Integrating over this energy density will give a logarithmic dependence on $D$. This logarithmic dependence on $D$ is not included in the analytic model~\eqref{Rcol}, \eqref{rkeq}.

The spin exchange energy in Fig.~\ref{stretchE} increases as a function of the vortex stretching $r$. This means that the $\psi_{\pm 1}$ vortices of a PCV are confined i.e. they cannot form an unbound state. Fig.~\ref{vdata}(b) suggests that as the $\psi_{\pm 1}$ vortices separate, spin waves carry away energy so that the stretching is reduced. However there is still an average increase in stretching as the PCVs collide. If this stretching continues to increase, the spontaneous creation of a PCV vortex antivortex pair will become favourable so that the spin exchange energy is reduced. This is analogous to color confinement in quantum chromodynamics~\cite{greensite2011}. To estimate the stretching required to observe this effect, we note that for large stretching Fig.~\ref{stretchE} predicts that the spin-exchange energy will scale as $\left(\hbar^2 n_0/M\right) r/\xi_s$. The energy required to spontaneously create a PCV vortex antivortex pair scales as $\left(\hbar^2 n_0/M\right)\ln\left(D/\xi_s\right)$. For $r/\xi_s\gtrsim\ln \left(D/\xi_s\right)$, the spontaneous creation of a PCV vortex antivortex pair could lower the system's free energy. The stretching may reach a maximum distance, however, beyond which any excess energy goes into spin wave production~\cite{turner2009}. It is possible that this could occur before the spontaneous creation of a PCV vortex antivortex pair. Reaching this large stretching regime for two PCVs in a setup like in Fig.~\ref{vIms} would require a large initial PCV separation so that the vortex stretching can become sufficiently large. Starting the dynamics with a non-zero stretching could allow more modest initial PCV separations to be used, although it may be hard to stabilise such an initial condition against decay through spin wave production.

\section{Conclusion}\label{concsec}
In this paper we have applied a variational Lagrangian approach to derive the model of PCV dynamics introduced by Turner~\cite{turner2009}. We compared this model to simulations of a PCV vortex antivortex pair and find semi-quantitative agreement. The distinguishing feature of the PCV dynamics is a vortex mass, which arises from interaction effects within the PCV core. The Turner model can be obtained by approximating this interaction as a quadratic potential which confines the $\psi_{\pm 1}$ components of a PCV. However, our numerics reveal higher order dynamic effects, which we suspect arise from core interactions coupling the PCV dynamics to spin waves. For high PCV energies, the nature of this coupling to spin waves may have a substantial effect on the confinement of the $\psi_{\pm 1}$ components. 
The Lagrangian formulation presented here paves the way for a more detailed study of the PCV core structure and dynamics, which can then be used to extend the Turner model. For example, ansatzes for density profiles of the spin components within the core could be included, as has been done for \textit{half-quantum} spin vortices~\cite{eto2011,kasamatsu2016}. The PCV mass depends on the quadratic Zeeman energy $q$, both explicitly in Eq.~\eqref{mdef} and implicitly through the parameter $a$ from Eq.~\eqref{fsimp}. This opens up the possibility of exploring PCV dynamics with a spatially dependent $q$, which could be done using the Lagrangian formulation presented here. It would also be interesting to explore the effects of a non-zero net $F_z$ magnetization on the PCV dynamics.

We have restricted our study to uniform systems for the purpose of characterising the interaction between PCVs. It is known that a scalar vortex in a harmonic trap can exhibit a precession around the trap centre~\cite{rokhsar1997}. It would be interesting to consider the effects of an harmonic trap on the dynamics of a PCV, where not only the density but also the spin properties (through $q_0$) vary spatially.

It should be possible to observe the dynamics of PCVs in current experiments using magnetization sensitive imaging (e.g.~see \cite{Higbie2005a,Sadler2006a}). Indeed, recent experiments in antiferromagnetic spin-1 condensates \cite{seo2015,seo2016} have been able to prepare half-quantum spin vortices and monitor their subsequent evolution and annihilation. For the ferromagnetic system PCVs can be spontaneously generated in a low-temperature quench from the polar phase to easy plane phase (see \cite{Sadler2006a,Damski2007a,Lamacraft2007a,Saito2007b} and Fig.~\ref{phasediag}), where the evolution of these vortices then determines the easy plane phase ordering dynamics~\cite{williamson2016a,williamson2016b}.

\section*{Acknowledgments}
LAW acknowledges valuable discussions with Ari Turner and Matt Reeves.
We gratefully acknowledge support from the Marsden Fund of the Royal Society of New Zealand.

\end{document}